\documentclass[12pt]{article}
\usepackage{amsmath,amssymb,amsfonts} % Typical maths resource packages
\usepackage{graphics}                 % Packages to allow inclusion of graphics
\usepackage{color}                    % For creating coloured text and background
\usepackage[cp1252]{inputenc}         %para respetar acentos en el momento de compilar
\usepackage[T1]{fontenc}
\usepackage{epsfig}

\begin{document}
\input epsf \renewcommand{\topfraction}{0.8}
\pagestyle{empty} \vspace*{5mm}
\begin{center}
\Large{\bf Brane-Worlds at the LHC:\\ Branons and KK-gravitons
} \\
\vspace*{1cm} \large{ Jose A. R. Cembranos, Rafael L. Delgado and Antonio Dobado
\\} \vspace{0.5cm}
 {\it Departamento de  F\'{\i}sica Te\'orica I,\\
 Universidad Complutense de
  Madrid,\\ 28040 Madrid, Spain}

\vspace{0.2cm} \vspace*{0.6cm} {\bf ABSTRACT} \\ \end{center}

We study the possibility of testing some generic
properties of Brane-World scenarios at the LHC. In particular, we
pay attention to KK-graviton and branon production. Both signals can
be dominant depending on the value of the brane tension. We
analyze the differences between these two signatures.
Finally, we use recent data in the single photon channel from the
ATLAS collaboration to constraint the parameter space of both
phenomenologies.

%\pacs{95.35.+d, 11.10.Kk, 11.25.Wx}
%pacs 95.35.+d, 11.10.Kk, 11.25.Wx

\section{Inroduction}

All observations carried out so far confirm the fact that there are three spatial dimensions.
There is no experimental evidence that points to the existence of additional dimensions.
However, there are numerous extensions of the Standard Model (SM) of particles that postulate
the existence of such dimensions due to different theoretical reasons, such as super gravity or
string theory (read \cite{Rubakov} for different reviews of the subject).
In the end of the past century, it was suggested that in particular constructions
associated to these models, the SM particles could be understood as confined fields into three
spatial dimensional manifolds or branes. On the contrary, the gravitational interaction has access
to the total or bulk space. In this scenario, the fundamental scale of gravitation is not the Planck scale
$M_P$,  but another different scale $M_D$ that can be much lower \cite{Rubakov:1983bb,ADD}.

This proposal, known in the literature as Brane World, opened a new range of theoretical approaches
and experimental possibilities to test the existence of new spatial dimensions. On the one hand, the
size of the extra dimensions is much less constrained than in the old Kaluza-Klein (KK) theories.
On the other hand, the lower gravitational fundamental scale allows to analyze the hierarchy problem
from a completely different perspective. Finally, the aspect that has made these
as attractive models and explains the large number of papers appeared in the last years, is the rich
phenomenology presented in accessible sensitivity ranges to present or future experiments.

The existence of extra dimensions leads to new degrees of freedom. The propagating gravitons along the
additional space develop a tower of Kaluza-Klein (KK) excitations from the four dimensional point of view.
On the other hand, the presence of the brane gives rise to the existence of another type of fields.
These models predict the existence of branons, particles associated to fluctuations of the brane
in the extra dimensions. The phenomenology associated with these two types of
new particles has been studied in different works. Specifically, these studies
have focused on potential signatures at particle accelerators (through real \cite{Coll} or virtual \cite{BWRad} processes),
astrophysical \cite{astro,indirect} and cosmological \cite{cosmo} observations. The study of KK-gravitons allows to constraint the number and size of extra dimensions under different assumptions. The analysis of branons restricts fundamental features of brane (such as its tension) an local properties of the bulk space.

\section{Settings of the Brane-World scenario}

The study of gravitational phenomena at the LHC is well established under the assumption of extra dimensions.
In particular, one of the most popular
possibilities is the so called Brane-World scenario (BWS). The original idea was proposed by Rubakov and Shaposhnikov
\cite{Rubakov:1983bb}, but more recently Arkani-Hamed,
Dimopoulos and Dvali, and also Antoniadis \cite{ADD}, introduced the so
called ADD scenario where the SM fields (or any suitable extension
of it) are confined (through some unspecified mechanism) to live
in a $3$ dimensional brane (the World brane) while the
gravitational field lives in the whole $D$ dimensional bulk space.
The extra dimensions are assumed to be compact and the World Brane
has a tension $\tau=f^4$ ($f$ is the brane tension parameter).
Its thickness depends on the underlying physics producing the brane but at relatively low energies it can be safely neglected. Thus the
main idea of the Brane-World scenario  is to assume that our usual $1 + 3$ dimensional world is some sort of three dimensional
 object (the brane) living in a higher $D$ dimensional bulk space $ {\cal M}_D$ with  $d$ additional spacial dimensions so that $D=4+d$.

In order to introduce some important concepts that we will be using
later, we will split the $D$ manifold as
\begin{equation}
{\cal M}_D={\cal M}_4 \times K_d,
\end{equation}
where ${\cal M}_D$ is called the bulk space and ${\cal M}_4$ is the
standard $1+3$ dimensional space-time brane manifold. In particular, we
can take ${\cal M}_4$ to be the $4$-dimensional
Minkowski space. The extra dimension space $K_d$ will be
assumed to be compact, which in particular means that it is a finite
volume manifold. Now we introduce the coordinates
$X^M=(x^\mu,y^m)$ where $x^\mu$ parametrizes  ${\cal M}_4$
($\mu=0,1,2,3$) and $y^m$ parametrizes  $K_d$ ($m=1,2,...,d$).
Also we choose the bulk-space metric $G_{MN}$ with signature $(+,-,-,-...)$ so that
\begin{equation}
ds^2=G_{MN}dX^MdX^N=g_{\mu\nu}dx^{\mu}
dx^{\nu}-\gamma_{mn}dy^mdy^n,
\end{equation}
where the metric $\gamma$ is positive definite and, according
to the compactness of $K_d$,
\begin{equation}
V_K=V(K_d)=\int d^dy\sqrt{|\gamma|} < \infty.
\end{equation}
For simplicity, we now consider a free real scalar field $\Phi$ of mass $M$ propagating
in the bulk with action
\begin{equation}
S[\Phi,G]=\int_{{\cal M}_D}d^DX\sqrt{|G|}[\frac{1}{2}(\partial_M
\Phi)^2-\frac{1}{2}M^2\Phi^ 2].
\end{equation}
The corresponding Euler-Lagrangian (Klein-Gordon) equation is
\begin{equation}
(-\Box_D -M^2)\Phi=0,
\end{equation}
where the $D$ dimensional d'Alambert operator is defined as
\begin{equation}
\Box_D \Phi=\nabla_M\nabla^M
\Phi=\frac{1}{\sqrt{|G|}}\partial^M(\sqrt{|G|}\partial_M\Phi).
\end{equation}
The d'Alambert operator can be written as
\begin{equation}
\Box_D =\Box_4-\Box _d,
\end{equation}
where obviously $\Box_d$ is the d'Alambert operator on $K_d$ and
it is positive (i.e. it has positive eigenvalues). Introducing a complete set of
orthonormal functions $Y_n=Y_n(y)$ on $K_d$, which are solutions of
the Klein-Gordon equation
\begin{equation}
\Box_dY_n(y)=\lambda_n Y_n(y),
\end{equation}
we have that, due to the compactness of $K_d$, the spectrum ${\lambda_n }$
is discrete. In addition, we can normalize the $Y_n(y)$ functions so that
\begin{equation}
\int_{{\cal K}_d}d^d y\sqrt{|\gamma|}Y_n(y)^*Y_m(y)=\delta_{nm}.
\end{equation}
Probably the simplest example of this setting is the $d$ dimensional torus
$K_d=T_d=S_1 \times S_1\times ...\times S_1$ ($d$ times). On each
circle $S_1$ we introduce the coordinate  $y^m$ which clearly is
periodic in the sense that $y^m$ and $y^m+2 \pi R$ represent the
same point ($R$ is the common radius of the circles). Thus we can
consider only $y^m$ values lying in the interval $y^m \in [0,2\pi
R]$. The volume of the torus is $V_T=V(T_d)=(2\pi R)^d$ and
$\Box_d=\nabla_d^ 2$ is just the $d$  dimensional Laplace
operator. Its eigenvalues $\lambda_n$ can be labeled by a set of
$d$ integers $n=(n_1,n_2,...,n_d)$ and they can be easily found to be
\begin{equation}
\lambda_n=\lambda_{(n_1,n_2,...,n_d)}=\sum_{m=1}^d\frac{n_m^2}{R^
2},
\end{equation}
and the corresponding eigenfunctions are
\begin{equation}
Y_{(n_1,n_2,...,n_d)}(y)=\frac{1}{\sqrt{V_T}}\exp
\{\frac{i\sum_{m=1}^d n_m y^m}{R}\}.
\end{equation}
Notice that $\lambda_0=0$ ($Y_0=1/\sqrt{V_T}$) but, for $n\neq 0$,
$\lambda_n$ goes as $1/R^ 2$.

In the general case we can expand the bulk field $\Phi$ in terms
of the $Y_n(y)$ as follows:
\begin{equation}
\Phi(X)=\Phi(x,y)=\sum_n \phi_n(x) Y_n(y).
\end{equation}
This is the so called Kaluza-Klein (KK) expansion, and plugging it
in the $\Phi$ action one gets
\begin{eqnarray}
S[\Phi,G]=\int_{M_D}d^4x\sqrt{|g|}  \{
\phi_0(-\Box_4-M^2)\phi_0  \\ \nonumber
 +\sum_{n\neq 0}\phi_n^*(-\Box_4-M_n^2)\phi \},
\end{eqnarray}
where
\begin{equation}
M_n^ 2=M^ 2+\lambda_n.
\end{equation}
For the case of the torus we have
\begin{equation}
\Phi(x,y)=\frac{1}{(2\pi R)^{d/2} }\sum_{(n_1,n_2,...,n_d)}
\phi_{(n_1,n_2,...,n_d)}(x)\exp \{\frac{i\sum_{m=1}^{d} n_m
y^m}{R}\},
\end{equation}
where $\phi_{(n_1,n_2,...,n_d)}(x)$ are the KK modes for the
$\Phi$ field. They can be considered $M_4$ fields with masses:
\begin{equation}
M_{(n_1,n_2,...,n_d)}^ 2=M^ 2+\frac{\sum_{m=1}^d n_m^2 }{R^2}
\end{equation}
For the simple case $d=1$ and $M=0$,
$\phi_0$ becomes a real zero mode. Notice also that for $n\neq 0$
the KK modes are complex and massive even when $M^ 2=0$ (no mass
term on the bulk for the $\Phi$ field). Also since $\Phi$ is a real
field as $Y_n^*(y)=Y_{-n}(y)$ we have $\phi_n^*(y)=\phi_{-n}(y)$.
Thus the complex conjugate of some KK mode represent the same mode
propagating in the opposite direction of the $T_d$ internal space.

In the general case a real bulk scalar  field $\Phi(X)$ is
equivalent to a KK tower of massive complex $M_4$ fields
$\phi_n(x)$ with masses $M^2_n=M^ 2+\lambda_n$.  At low energies $E<<
1/R$ ($R$ being the typical size of the extra dimensions i.e. $V_K
\sim R^ d$) only the real zero mode survives. However, at higher
energies more and more KK modes become relevant and must be taken
into account.

Note that even if the size of the extra dimensions is too small to
be directly observable, their existence could be probed by
detecting these KK modes of the effective four dimensional theory.
The spectrum of the KK tower will give us information about the
geometry of the internal space. At low energies only zero modes
can be excited (dimensional reduction).

\section{Gravitons}

According to the general idea that  our universe is a $3$ brane, i.e. a
$3$-dimensional smooth object living in a higher dimensional space
(the $D$ dimensional bulk space) and  using the
notation introduced above it is very easy to find that
\begin{equation}
M_P^ 2=M_4^ 2=V(K_d)M_D^{D-2},
\end{equation}
where $M_D$ is the $D$ dimensional Plank scale defined. Thus, the Einstein-Hilbert action is given by
\begin{equation}
S_{EH} =  \frac{M_D^{D-2}}{16 \pi}\int_{{\cal M}_D}d^D
X\sqrt{|G|}[R_D-(D-2)\Lambda_D],
\end{equation}
where we have introduced the bulk cosmological constant $\Lambda_D$. Obviously, $M_P=M_4$
(we are using $\hbar = c = 1$ units), so that $M_{P} \simeq1.2 \times 10^
{19}\,{\rm GeV}$.

One of the important points of original idea of the ADD scenario is to have $R$
large enough so that the $D$ dimensional Plank mass $M_D$ (the
fundamental scale of gravity) could be of the order of the $TeV$ scale,
thus solving, or at least putting in a completely new setting, the
hierarchy problem since for $v\simeq 250\,{\rm GeV}$ being the
electroweak symmetry breaking scale we could have
\begin{equation}
M_D \sim 4 \pi v \ll M_P.
\end{equation}
Therefore, the huge hierarchy is produced by the {\it large} volume
of the extra dimension space $K_d$. For example for $d=1$,
$M_D\sim 1\,{\rm TeV}$ requires $R \sim 10^ {13}\,{\rm cm}$, which is ruled out by
our knowledge of the Newton law at the Solar System scale. For
$d=2$, $R \sim 0.1\,{\rm mm}$, which is close to the experimental limit
coming from the study of possible deviations from the Newton law
at the sub millimeter scale. For $d \geq 3$, $R$ must be of the
order or smaller than $10^ {-7}\,{\rm cm}$, which in principle is well
below any experimental constraint.

The simplest action describing the ADD model is
\begin{equation}
S_{ADD} =  \frac{M_D^{D-2}}{16 \pi}\int_{{\cal M}_D}d^D
X\sqrt{|G|}[R_D-(D-2)\Lambda_D]+\int_{ {\cal M}_4}d^4 x\sqrt{|g|}(
{\cal L}_{SM}(g,\Phi)-\tau ),
\end{equation}
where ${\cal M}_4$ is the brane world-sheet with coordinates $x^
\mu$, ${\cal L}_{SM}$ is the SM Lagrangian defined on ${\cal M}_4$, $\Phi$
represents all the SM fields and the last term is just the
Nambu-Goto action for the brane. Any point on ${\cal M}_4$ will
have bulk coordinates $Y^M=Y^M(x)$. Then the interval on this $4$
dimensional manifold is given by
\begin{equation}
ds^ 2 =G_{MN}dY^MdY^N=G_{MN}\frac{\partial Y^M}{\partial
x^\mu}\frac{\partial Y^N}{\partial x^\nu}d x^\mu d x^\nu \equiv
g_{\mu\nu} dx^\mu dx^\nu.
\end{equation}
The $g_{\mu\nu}$ metric defined on ${\cal M}_4$ is called the
induced metric (or the $G_{MN}$ pull-back on ${\cal M}_4$). It
includes the curvature coming from the bulk ${\cal M}_D$ and the
own ${\cal M}_4$ curvature coming from the different ways in which
${\cal M}_4$ can live in ${\cal M}_D$.

In order to study some of the properties of this model, and for
other reason that will be explained at the end of this section, we
will concentrate mainly on the simple case ${\cal M}_D=M_4 \times
T_d$. Notice that in particular this means that we are neglecting
any brane fluctuation so this can be understood as a case of having
the brane tension scale $f$ much larger than the other relevant
scales in the system. In the next section we will consider the
effects produced by these brane fluctuations for lower
 values of the tension parameter $f$ (flexible brane case).

Now, in order to study the graviton excitations, we write the bulk
metric as
\begin{equation}
G_{MN}(x,y)=\eta_{MN}+\frac{2 }{\bar M_D^{1+d/2}}h_{MN}(x,y),
\end{equation}
where $\bar M_D^{D-2}\equiv M_D^{D-2}/ 4 \pi$ is the reduced
fundamental scale and $h_{MN}(x,y)$ is the bulk graviton field.
The normalization is chosen so that the corresponding action at
the lowest order is the canonical one
\begin{eqnarray}
S[h]=\int_{{\cal M}_D}d^D X[\frac{1}{4}\partial^R h^{MN}
\partial_R h_{MN}-\frac{1}{2}\partial^R h^{MN}
\partial_M h_{RN}  \nonumber      \\
+\frac{1}{2}\partial^M h
\partial^L h_{LM}-\frac{1}{4}\partial^M h
\partial_M h],
\end{eqnarray}
where $h=\eta_{MN}h^{MN}$. The graviton field can be KK expanded
as
\begin{equation}
h_{MN}(x,y)=\frac{1}{(2\pi R)^{d/2} }\sum_{(n_1,n_2,...,n_d)}
h_{MN}^{(n_1,n_2,...,n_d)}(x)\exp \{\frac{i\sum_{m=1}^{d} n_m
y^m}{R}\},
\end{equation}
where $h_{MN}^{(n_1,n_2,...,n_d)}(x)$ are the KK modes for the
graviton field with masses
\begin{equation}
M_{(n_1,n_2,...,n_d)}^ 2=\frac{\sum_{m=1}^d n_m^2 }{R^2}.
\end{equation}
Therefore, in addition to the usual massless graviton, we will have
an infinite tower of complex massive gravitons. One important
observation here is that the gap or mass distance between two
consecutive massive gravitons goes as $\triangle M \sim 1/ R$. This
means that for large enough extra dimensions the KK graviton
spectrum can be considered as almost continuous. As we will see
later this is an important fact that opens the possibility of
producing gravitons in  a detectable rate under some conditions.

Thus, the massless zero mode graviton
$h_{\mu\nu}(x)$ has a whole tower of massive KK partners,
$h^{n}_{\mu\nu}(x)$  which  are massive  $J=2$ fields with $5$
 physical polarization states. The additional degrees of freedom come
from a sort of Higgs mechanism, present in Kaluza-Klein theories,
where  the field $ h^{n}_{\mu\nu}(x)$ {\it eats} some of the extra
dimensional excitations producing the tower of massive $J=2$ KK
modes. The effective Lagrangian describing the free evolution of
these massive fields can be taken to be the well known Fierz-Pauli
action
\begin{eqnarray}
S_{FP}[h]=\sum_{n}\int_{ M_4}d^4 x[\frac{1}{4}\partial^{\rho}
h^{\mu\nu(n)}
\partial_{\rho} h_{\mu\nu}^{(n)}-\frac{1}{2}\partial^{\rho} h^{\mu\nu(n)}
\partial_{\mu} h_{\rho\nu}^{(n)} +\frac{1}{2}\partial^{\mu} h^{(n)}  \partial^{\lambda}
 h_{\lambda\mu}^{(n)}  \nonumber\\
-\frac{1}{4}\partial^{\mu} h^{(n)}\partial_{\mu}
h^{(n)}-\frac{1}{4}M^2_n(h^{\mu\nu(n)}h_{\mu\nu}^{(n)}-(h^{(n)})^2)],
\end{eqnarray}
where, for example, $n$ should be understood as $n=(n_1,n_2,...,n_d)$
in the torus case. In particular, $M^2_n=M_{(n_1,n_2,...,n_d)}^ 2$.
The Above action naturally leads to the set of equations
\begin{eqnarray}
           h^{(n)} & = & 0     \nonumber \\
              \partial_{\mu} h^{\mu\nu(n)}    &  =  &  0  \nonumber    \\
              (\Box + M^2_n)h^{\mu\nu(n)}  & = & 0.
\end{eqnarray}
Here the first two equations are the five constraints that
reduce the original  degrees of freedom of the symmetric tensors
$h^{\mu\nu(n)}$ from ten to five and the last one is just the
Klein-Gordon equation expected for free massive bosons. Now, in
order to study the interaction between massive gravitons and the SM
particles we start from the SM piece of the ADD action
\begin{equation}
S_{SM}[g,\Phi] = \int_{ {\cal M}_4}d^4 x\sqrt{|g|}{\cal
L}_{SM}(g,\Phi),
\end{equation}
and then we expand it around the $\eta_{\mu\nu}$ Minkoskian ($M_4$) brane metrics
\begin{equation}
S_{SM}[g,\Phi] =\int_{M_4}d^4 x {\cal L}_{SM}(\eta,\Phi)+
\int_{M_4}d^4 x  \frac{  \delta S_{SM} }{\delta g_{\mu\nu}(x) }
\mid _{g=\eta}\delta g_{\mu\nu}(x)+ ...
\end{equation}
But in our setting we have
\begin{equation}
\delta g_{\mu\nu}(x)=\frac{2 }{\bar M_D^{1+d/2}}h_{\mu\nu}(x),
\end{equation}
and the SM energy momentum tensor is
\begin{equation}
T^{\mu\nu}_{SM}= -\frac{2}{\sqrt{|g|}} \frac{ \delta S_{SM}
}{\delta g_{\mu\nu}(x) } \mid _{g=\eta}.
\end{equation}
Therefore, the $S_{SM}$ action can be split as the usual SM
action in flat space-time $M_4$ plus an interacting term $S_{int}$
given by
\begin{equation}
S_{int}[h,\Phi] =      -\frac{1}{\bar M_D^{1+d/2}} \int_{M_4}d^4 x
T^{\mu\nu}_{SM} h_{\mu\nu}.
\end{equation}
Or using the relation between the $D$ dimensional fundamental
scale of gravity $M_D$, the Plank scale $M_P$ and the KK mode
expansion for the graviton fields
\begin{equation}
S_{int}[h,\Phi] =      -\frac{1}{\bar M_P} \sum_{(n_1,n_2,...,n_d)}
\int_{M_4}d^4 x  T^{\mu\nu}_{SM} h_{\mu\nu}^{(n_1,n_2,...,n_d)},
\end{equation}
according to the expectation that SM and graviton interactions are
suppressed by the Planck mass.

From this action, following the standard procedure,  it is
possible to obtain the Feynman rules for the different couplings
such as graviton-fermion antifermion, graviton-photon-photon,
graviton-photon-fermion-antifermion, graviton-gluon-gluon-gluon
and many others. Some attention must be paid to the gauge fixing
conditions for the graviton field, which should give rise to the
appropriate propagators  and polarization wave functions, that must
reproduce the two polarization states of the massless graviton and
the five polarization states of the massive gravitons.
In the case of virtual gravitons one should
also pay attention to the corresponding ghost fields.

Thus, it is possible for instance to compute the
amplitude of the process $e^+e^- \rightarrow \gamma h^{n}$.
The signal for this reaction would  be very clear since gravitons
escape from detection and then we are left just with one single
photon event with missing energy and $P_T$. Nevertheless the
cross-section for producing one graviton is strongly suppressed by
the Planck mass and one expects
\begin{equation}
\sigma \sim \frac{1}{M_P^ 2}.
\end{equation}
However, if one considers the cross section for producing {\it any}
KK graviton, things are completely different. As it was commented
above for large $R$, the KK spectrum can be considered continuous.
Let us define $N(k)$ as the number of KK modes with modulus $\lvert
\vec {k}\rvert$ of the extra dimension momentum $\vec
{k}=(k_1,...,k_d)$ lesser or equal than $k$. Then, it is easy to see
that
\begin{equation}
dN \sim R^dS_{d-1}k^ {d-1}dk,
\end{equation}
where
\begin{equation}
S_{n}=\frac{(2\pi)^{n/2}}{\Gamma(n/2)}
\end{equation}
Therefore, for some given energy $E$, the number of available KK
gravitons is
\begin{equation}
N(E)=\int_0^ {E}n(E')dE' \sim \frac{S_{d-1}M_P^2E^d}{dM_D^ {d+2}},
\end{equation}
where $n(E)=dN/dE$ is the KK states energy density.
Thus, we finally arrive to the conclusion that the cross section for
producing {\it any} KK graviton goes as \cite{GRW}
\begin{equation}
\sigma \sim \frac{S_{d-1}}{d}\frac{E^d}{M_D^ {d+2}}
\end{equation}
and therefore it is not suppressed by the Planck mass but by
the fundamental scale $M_D$ which in this framework is supposed to
be of the order of the $TeV$. This is a very interesting property
of the ADD model since it opens the possibility of having
detectable gravitational interactions at the LHC.

From this example based in the case $K_d=T_d$ we see that total cross section for producing gravitons in the large $R$ limit (which means continuous spectrum) depends on the KK state density $n(E)$ i.e. on the spectrum of the KK gravitons.
Thus in principle by measuring carefully the cross section for graviton production, let us say, at the LHC, it could be possible to obtain information about the spectrum which in turns could carry information about the $K_d$ (extra dimension space) geometry and topology. We can illustrate this idea by comparison between the two simple $d=2$ cases $K_2=T^2$ and $K_2=S^2$.
As discussed above the graviton spectrum for the first case is given by
\begin{equation}
M^2_{(n_1,n_2)}=\frac{1}{R^2}(n^2_1+n^2_2).
\end{equation}
In the $S^2$ case the volume of the extra dimension space is given by $V(S^2)=4 \pi R^2$ where $R$ is the sphere radius  (note the different geometrical meaning of $R$ in both cases).  The graviton field can be KK expanded
as
\begin{equation}
h_{MN}(x,y)=\frac{1}{ R }\sum_{l,m}
h_{MN}^{lm}(x)Y_{lm}(y),
\end{equation}
where $Y_{lm}(y)$ with $y=(\theta,\phi)$ are the standard spherical harmonics
and thus $l=0,1,2...$ and $m=-l,...0,...l$ and the corresponding spectrum is
\begin{equation}
M^2_{lm}=\frac{1}{R^2}l(l+1).
 \end{equation}
Therefore, as expected, the two spectra are different having different gaps and degeneracies. In the large $R$ limit we can obtain the state densities. According to our previous discussion for the $T^2$ case, we get
\begin{equation}
n_{T^2}(E)=\frac{\pi}{2}R^2E.
\end{equation}
For the $S^2$ case it is not difficult to find
\begin{equation}
n_{S^2}(E)=2R^2E.
\end{equation}
This example shows that in the continuous limit, both the gap and degeneracy information
contained in the discrete spectra, are washed out, being the two cases indistinguishable,
at least from the practical point of view, as concerned to the graviton production cross-sections.
For this reason, we will concentrate on the torus case,
as representative of many other possible compactified extra dimension spaces.

\section{Brane fluctuations (branons)}

In this section we consider branons \cite{sun,DoMa},
another kind of excitations that could possibly be present in the Brane-World
scenarios which are particularly interesting when the scale tension parameter $f$
is low enough, i.e. in the case of a flexible brane. As in the previous discussion the brane lies
along ${\cal M}_4$ but to start with we will neglect the gravitons (see next section).
The bulk space ${\cal M}_D$ metrics will be assumed to have the general form \cite{DoMa,BW1}
\begin{eqnarray}
 G_{MN}&=&
\left(
\begin{array}{cccc}
\tilde g_{\mu\nu}(x)&0\\ 0&-\tilde g'_{mn}(y)
\end{array}\right).
\label{bulkmetric}
\end{eqnarray}

The position of the brane in the bulk can be parametrized as
$Y^M=(x^\mu, Y^m(x))$
where we have chosen the bulk coordinates
so that the first four are identified with the space-time brane
coordinates $x^\mu$. We assume the brane to be created at a
certain point in $K_d$, i.e. $Y^m(x)=Y^m_0$, which corresponds to its
ground state. The induced metric on the brane in this particular case is
given by $g_{\mu\nu}=\tilde g_{\mu\nu}=G_{\mu\nu}$. However,
when brane excitations  are present, the induced metric is given by
\begin{eqnarray}
g_{\mu\nu} & = & \partial_\mu Y^M\partial_\nu Y^N G_{MN}(x,Y(x))
\nonumber     \\
& = & \tilde
g_{\mu\nu}(x,Y(x))-\partial_{\mu}Y^m\partial_{\nu}Y^n\tilde
g'_{mn}(Y(x)).
\end{eqnarray}

Since the mechanism responsible for the creation of the brane is
in principle unknown, we will assume that the brane dynamics can
be described by an effective action.  At low energies the dominant term
is the one, having the appropriate symmetries, with the least possible
number of derivatives
of the induced metric. This principle leads us to
\begin{equation}
S_B=-f^4\int_{M_4}d^4x\sqrt{g},
\label{Nambu4}
\end{equation}
where $d^4x\sqrt{g}$ is the volume element of the brane. Notice
that this lowest order term is the Nambu-Goto action introduced below.

In the absence of the 3-brane, the  metric (\ref{bulkmetric})
possesses an isometry group which we will assume to be of the form
$G({\cal M}_D)=G({\cal M}_4)\times G(K_d)$. The presence of the brane will break spontaneously
all the $K_d$ isometries, except those that leave the point $Y_0$ (the brane ground state)
unchanged. The group $G(K_d)$ is spontaneously broken
down to $H(Y_0)$, where $H(Y_0)$ denotes the isotropy group (or
little group) of the point $Y_0$ and we can define the
coset space $K=G({\cal M}_D)/(G({\cal M}_4)\times H(Y_0))=G(K_d)/H(Y_0)$.

When  the $K_d$ space is homogeneous the little group $H(Y_0)$ is
$Y_0$ independent and $H(Y_0) \equiv H$. The coset $K$ is isomorphic
to $K_d$ and the isometries are just translations. In this case the branon fields
($\pi$), defined as Gaussian coordinates on the coset $K$, can be identified,
with properly chosen coordinates in the extra
space $K_d$, as for example
\begin{equation}
\pi^{\alpha}=f^2\delta^{\alpha}_m y^m.
\end{equation}
In the following, for the sake of simplicity, we will consider only $K_d$ homogeneous spaces.

According to the previous discussion, we can write the induced
metric on the brane in terms of branon fields as
\begin{equation}
g_{\mu\nu}=\tilde g_{\mu\nu}(x)- \tilde
g'_{mn}\frac{\partial Y^m}{\partial\pi^\alpha}\frac{\partial
Y^n}{\partial\pi^\beta}\partial_{\mu}\pi^\alpha
\partial_{\nu}\pi^\beta.
\end{equation}

Introducing the metrics $h_{\alpha\beta}(\pi)$  as
\begin{equation}
h_{\alpha\beta}(\pi)=f^4 \tilde g'_{mn}(Y(\pi))\frac{\partial
Y^m}{\partial\pi^\alpha}\frac{\partial Y^n}{\partial\pi^\beta},
\end{equation}
we have
\begin{eqnarray}
g_{\mu\nu}&=&\tilde
g_{\mu\nu}(x)-\frac{1}{f^4}h_{\alpha\beta}(\pi)\partial_{\mu}\pi^\alpha
\partial_{\nu}\pi^\beta.
\label{induced}
\end{eqnarray}

The above scheme leading to massless branons is only valid
if the isometry pattern introduced
before is exact. However, in more general situations, these symmetries are only approximately realized and
branons will acquire  mass \cite{BW1,mass}. In order to illustrate how this could happen
explicitly, let us perturb the four-dimensional components of the
background metric and let $\tilde g_{\mu\nu}$ be dependent, not
only on the $x$ coordinates, but also on the $y$ ones \cite{BW1,mass},
\begin{eqnarray}
 G_{MN}&=&
\left(
\begin{array}{cccc}
\tilde g_{\mu\nu}(x,y)&0\\ 0&-\tilde g'_{mn}(y)
\end{array}\right).
\end{eqnarray}
This has to be done in such a way
that the $G(K_d)$ piece of the full isometry group is explicitly
broken. Notice that the breaking of the $G(K_d)$ group by perturbing
only the internal
metric $\tilde g_{mn}'(y)$  does not lead to a mass term for the
branons.

In order to calculate the branon mass matrix, we need to know
first the  ground state around which the brane is fluctuating.
With that purpose, we will
consider for simplicity the lowest-order action, given by
\begin{eqnarray}
S_{eff}^{(0)}[\pi]&=& -f^4
\int_{M_4}d^4x\sqrt{\tilde g(x,Y(x))},
\end{eqnarray}
which will have an extreme provided by
\begin{eqnarray}
\delta S_{eff}^{(0)}[\pi]=0\Rightarrow \delta\sqrt{\tilde
g}=\frac{1}{2}\sqrt{\tilde g}\tilde g^{\mu\nu}\delta \tilde
g_{\mu\nu}=0\Rightarrow \tilde g^{\mu\nu}\partial_{m} \tilde
g_{\mu\nu}=0.
\label{extremum}
\end{eqnarray}
This is a set of equations whose solution $Y^m_0(x)$ determines
the shape of the brane in its ground state for a given background
metric $\tilde g_{\mu\nu}$.
%And the used action is exact if $Y^m$ is constant: $Y^m=Y^m_0$.
In addition, the  condition for the energy to be minimum requires
\begin{eqnarray}
\left.\frac{\delta^2 S_{eff}^{(0)}}{\delta Y^m\delta
Y^n}\right|_{Y=Y_0} <0,
\end{eqnarray}
which means:
\begin{eqnarray}
\frac{f^4}{4}\sqrt{\tilde g}\tilde
g^{\mu\nu}(\partial_{n}\partial_{m} \tilde g_{\mu\nu}-2\tilde
g^{\rho\sigma}\partial_{n} \tilde g_{\nu\sigma}\partial_{m} \tilde
g_{\mu\rho}) > 0, \label{condsy}
\end{eqnarray}
i.e., the eigenvalues of the above matrix should be positive. This
implies that the action should have a minimum for static configurations.

In order to obtain the explicit expression of the branon
mass matrix, we expand  $\tilde g_{\mu\nu}(x,y)$ around
$y^m=Y^m_0$ in terms of the $\pi^\alpha$ fields:
\begin{eqnarray}
\tilde g_{\mu\nu}(x,y)&=&\tilde g_{\mu\nu}(x,Y_0)+\partial_m
\tilde g_{\mu\nu}(x,Y_0)(Y^m-Y^m_0)
\label{indbran} \\
&+&\frac{1}{2}\partial_m\partial_n\tilde
g_{\mu\nu}(x,Y_0)(Y^m-Y^m_0)(Y^n-Y^n_0)+...\nonumber \\ &=& \tilde
g_{\mu\nu}(x,Y_0)
+\frac{1}{f}V_{\alpha\mu\nu}^{(1)}\pi^\alpha
+\frac{1}{f^2}V_{\alpha\beta\mu\nu}^{(2)}\pi^\alpha\pi^\beta
+\dots\nonumber
\end{eqnarray}
%

%As we will explain below, we will discard the linear term in
%branon fields.
The linear term in branon fields is written as
\begin{eqnarray}
V_{\alpha\mu\nu}^{(1)}=\left.\partial_m\tilde
g_{\mu\nu}(x,y)\right\vert_{y=Y_0}\frac{\xi^m_\alpha}{kf},
\end{eqnarray}
where $\xi_\alpha$ are the Killing vectors corresponding to the broken generators
defining the coset $K \sim K_d$, i.e., those generators of
$G({\cal M}_D)=G({\cal M}_4)\times G(K_d)$ not present in $H$. These Killig vector are
normalized so that
$k^2=16\pi /M_P^2$, being $M_P$ the four-dimensional Planck mass.

The quadratic term takes the general form
\begin{eqnarray}
V_{\alpha\beta\mu\nu}^{(2)}=\frac{f^2}{2}\left.\partial_m
\tilde
g_{\mu\nu}(x,y)\right\vert_{y=Y_0}\left.\frac{\partial^2
Y^m}{\partial \pi^\alpha\partial \pi^\beta}\right\vert_{\pi=0}
+
\frac{1}{2}\left.\partial_m\partial_n\tilde
g_{\mu\nu}(x,y)\right\vert_{y=Y_0}\frac{\xi^m_\alpha\xi^n_\beta}{k^2f^2}.
\end{eqnarray}
Here, we have used the fact that the action of an
element of $G(K_d)$ on $K_d$ will map $Y_0$ into
some other point with coordinates
\begin{equation}
Y^m(x)=Y^m(Y_0,\pi^\alpha(x))=Y^m_0+\frac{1}{k
f^2}\xi^m_\alpha(Y_0)\pi^\alpha(x)+ O(\pi^2).
\end{equation}
Substituting the above expression back in
(\ref{induced}), we get the expansion of the induced metric in
branon fields:
\begin{equation}
g_{\mu\nu}=
\tilde g_{\mu\nu}(x,Y_0)-\frac{1}{f^4}\delta_{\alpha\beta}\partial_{\mu}\pi^\alpha
\partial_{\nu}\pi^\beta
+\frac{1}{f}V_{\alpha\mu\nu}^{(1)}\pi^\alpha
+\frac{1}{f^2}V_{\alpha\beta\mu\nu}^{(2)}\pi^\alpha\pi^\beta
+O (\pi^4).
\end{equation}

We have also used the fact that since $\pi^\alpha$ must be
properly normalized scalar fields, the $Y^m$ coordinates
should be normal and geodesic in a neighborhood of
$Y^m_0$ and, in particular, they cannot be angular coordinates.
This implies that we can write
$h_{\alpha\beta}(\pi=0)=\delta_{\alpha\beta}$.

Assuming for concreteness that, in the ground state, the
four-dimensional background metric is flat, i.e. $\tilde
g_{\mu\nu}(x,Y_0)=\eta_{\mu\nu}$, the appearance of the
$V_{\alpha_1\alpha_2...\alpha_i\mu\nu}^{(i)}$ tensors
in (\ref{indbran}) could break Lorentz invariance,
unless they factor out as
$V_{\alpha_1\alpha_2...\alpha_i\mu\nu}^{(i)}=
M_{\alpha_1\alpha_2...\alpha_i}^{(i)}\eta_{\mu\nu}/(4f^2)$.
With this assumption, the linear term
$V_{\alpha\mu\nu}^{(1)}$  vanishes identically
due to the condition of minimum for the brane energy (\ref{extremum}),
and the $M_{\alpha\beta}^{(2)}$ coefficient in the quadratic term
can be identified with the branon mass matrix.
Thus  we find
\begin{equation}
\sqrt{g}=1-\frac{1}{2f^4}\eta^{\mu\nu}\delta_{\alpha\beta}
\partial_{\mu}\pi^\alpha\partial_{\nu}\pi^\beta
+\frac{1}{2f^4}M_{\alpha\beta}^{(2)}\pi^\alpha\pi^\beta
+... \label{det}
\end{equation}
Notice that this expression requires that both $\partial \pi/f^2$
and $M^2\pi^2/f^4$ are small. This includes different types of
approaches, such as low-energy expansions with small branon masses
compared to $f$, or low-energy expansions with possible large
masses and small $\pi/f$ factors.

The different terms in the effective action can be organized according
to the number of branon fields,
\begin{equation}
S_{eff}[\pi]=S_{eff}^{(0)}[\pi]+ S_{eff}^{(2)}[\pi]+ ...\,,
\end{equation}
where the zeroth order term is just a constant. The free action
contains the terms with two branons,
\begin{eqnarray}
 S_{eff}^{(2)}[\pi]=\frac{1}{2}\int_{M_4}d^4x
(\delta_{\alpha\beta}\partial_{\mu}\pi^\alpha\partial^{\mu}\pi^\beta
-M^2_{\alpha\beta}\pi^\alpha\pi^\beta).
\end{eqnarray}

\vspace{.5cm}
In principle one can always diagonalize the squared mass matrix
$M^2_{\alpha\beta}$ to obtain the physical branon fields with masses $M_{\alpha}$.

In order to study the possible phenomenological consequences of the brane fluctuations it is very important to obtain the coupling of branons to the SM particles. To this end it is enough to consider the case
$\tilde g_{\mu\nu}=\eta_{\mu\nu}$. Now we  can proceed as in
\cite{DoMa}, where the SM action on the brane is expanded in
branon fields through the induced metric. Thus the
complete action, including terms up to two branons, is given by
\cite{BW1,Alcaraz}
\begin{eqnarray}
S_B&=& \int_{M_4}d^4x\sqrt{g}[-f^4+ {\cal L}_{SM}]
\nonumber\\
&=&\int_{M_4}d^4x\left[-f^4+ {\cal L}_{SM}(
\eta_{\mu\nu})  +
\frac{1}{2}\delta_{\alpha\beta}\partial_{\mu}\pi^\alpha
\partial^{\mu}\pi^\beta-\frac{1}{2}M^2_{\alpha\beta}\pi^\alpha\pi^\beta\right.
\nonumber\\
&+& \left.\frac{1}{8f^4}(4\delta_{\alpha\beta}\partial_{\mu}\pi^\alpha
\partial_{\nu}\pi^\beta-M^2_{\alpha\beta}\pi^\alpha\pi^\beta\eta_{\mu\nu})
T^{\mu\nu}_{SM} \right]
+{\cal O}(\pi^3),
\label{quadratic}
\end{eqnarray}
where $T^{\mu\nu}_{SM}(\eta_{\mu\nu})$ is the SM
energy-momentum tensor as defined in the case of gravitons considered in the previous section:

\begin{eqnarray}
T^{\mu\nu}_{SM}=-\left.\left(\tilde g^{\mu\nu}{\cal L}_{SM}
+2\frac{\delta
{\cal L}_{SM}}{\delta \tilde g_{\mu\nu}}\right)\right
\vert_{\tilde g_{\mu\nu}=\eta_{\mu\nu}}.
\end{eqnarray}
Notice that  no single branon
interactions, which will be related to
Lorentz invariance breaking, are present in this action.
In addition the quadratic
expression in (\ref{quadratic}) is valid for any $K_d$
space, regardless of the particular form of the metric
 $\tilde g'_{mn}$. In fact, the form of the couplings only depends
on the number of branon fields, their mass and the brane tension.
The dependence on the geometry of the extra dimensions will appear
only at higher orders, contrary to the case of gravitons.
Therefore branons interact always by pairs with the SM matter fields.
In addition, due to
their geometric origin, those interactions are very similar to the
gravitational ones since the $\pi$ fields couple to all
the matter fields through the energy-momentum tensor and with the
same strength, which is suppressed by a $f^4$ factor. In fact,
branons couple as gravitons do, with the identification
\cite{BW1,subs}:
\begin{eqnarray}
 -\frac{1}{\bar M_P} h_{\mu\nu}&\longrightarrow&\frac{1}{8f^4}
(4\delta_{\alpha\beta}\partial_{\mu}\pi^\alpha
\partial_{\nu}\pi^\beta-M^2_{\alpha\beta}\pi^\alpha\pi^\beta\eta_{\mu\nu}),
\end{eqnarray}
where $h_{\mu\nu}$ is the graviton field in linearized gravity.
As in the graviton case, by using standard methods, it is possible to find
the relevant Feynman rules, amplitudes and cross-section for producing
branons (by pairs) in, for example, the LHC. From the discussion above,
it is clear that these branons are in general massive, stable (at least the lightest of them)
and, therefore, they would scape to detection, being its main signature missing
energy and momentum as it is the case also of gravitons.

\begin{figure}[h]
\null\hfill\includegraphics[width=.8\textwidth]{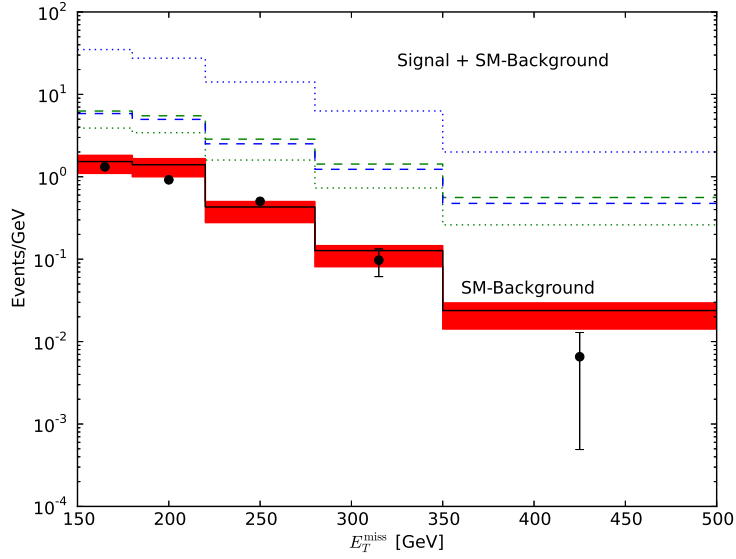}\hfill\null
\caption{The ATLAS $E_T^{\rm miss}$ distribution (black dots, $\sqrt{s}=7\,{\rm TeV}$, $\int L\,dt=4.6\,{\rm fb}^{-1}$) versus the SM background (red band, see~\cite{2012arXiv1209.4625T}) and our own computations for both the KK--Graviton+SM background (dashed lines) and the branon+SM background (dotted lines). The represented graviton models uses $M_D = 1\,{\rm TeV}$ and $N=2$ (lower blue dashed line), and $M_D = 1.5\,{\rm TeV}$ and $N=6$ (upper green dashed line). And the branon one, $M = 2\,{\rm TeV}$, $N=1$ and $f=60\,{\rm GeV}$ (upper blue dotted line) and $M = 1\,{\rm TeV}$, $N=1$ and $f=200\,{\rm GeV}$ (lower green dotted line).}\label{fig_MC}
\end{figure}

\section{Collider phenomenology: single photon channel}\label{collider}

From the discussion presented so far it is clear that probably the most outstanding property of
flexible Brane-World scenarios is the presence of
two types of generic excitations, namely: KK gravitons and branons.
Curiously enough the experimental signatures for producing these excitations
starting from SM particles in colliders as the LHC is in both cases
missing energy and transverse momentum. Therefore, finding an important number of missing energy events at the LHC, could be
suggesting some chances for a Brane-World case (note however that other scenarios like the MSSM could also produce such a signals, but in that case one expect the  production of many other new SUSY particles that should also be present).

The graviton production cross-section are of the order $\sigma_G \sim (ER)^d/M_P^2$ and the
branon production cross-section goes as $\sigma_B \sim E^6/f^8$. Therefore
it is clear that, for some given extra dimension space size $R$, the relative production rate of gravitons and branons is controlled by the brane tension parameter $f$. For rigid branes (high values of $f$) we expect graviton production dominance, while for flexible branes (low $f$) we expect branon production to be more abundant.

In the rest of this work we will study in detail the single photon channel for producing branons and KK-gravitons at the LHC and their characteristic missing energy and transverse momentum signatures. In the first case, we need
the cross-section of the subprocess $q \bar q
\rightarrow \gamma \pi\pi$, that was computed in \cite{BWHad}:
\begin{eqnarray}
&&\frac{d\sigma (q \bar q\rightarrow \gamma \pi \pi )}{dk^{2}dt}  \nonumber \\
&=&\frac{Q_{q}^{2}\alpha N(k^{2}-4M^{2})^{2}}{184320f^{8}\pi ^{2}\hat{s}%
^{3}tu}\sqrt{1-\frac{4M^{2}}{k^{2}}}(\hat{s}k^{2}+4tu)(2\hat{s}%
k^{2}+t^{2}+u^{2})\,,
\end{eqnarray}%
where $N$ is the number of branons (that we will assume degenerate and
equal to the number of extra dimensions),
$\hat s\equiv(p_1+p_2)^2$, $t \equiv(p_1-q)^2$,
 $u\equiv(p_2-q)^2$ and $k^2\equiv(k_1+k_2)^2$.
$p_1$ and $p_2$ are the initial quark and anti-quark four-momenta;
$q$, the final photon four-momentum; and $k=k_1+k_2$, the total branon
four-momentum. Thus, the contribution to the total cross section for the $p
p\rightarrow \gamma\pi\pi$ reaction is
\begin{eqnarray}
\sigma(p p\rightarrow \gamma\pi\pi)= \int_{x_{min}}^1
dx\int_{y_{min}}^1 dy \sum_q
 \bar q_{ p}(y;\hat s) q_{ p}(x;\hat s)   \nonumber\\
\int_{k^2_{min}}^{k^2_{max}}  dk^2 \int_{t_{min}}^{t_{max}}
dt\frac{d\sigma(q q \rightarrow \gamma\pi\pi)}{dk^2dt}.
\end{eqnarray}

Here $q_p(x;\hat s)$ and $\bar q_{ p}(y;\hat s)$ are the quark and anti-quark
distribution functions of the proton, $x$
and $y$ are the fractions of the protons energy carried by the
initial quark and anti-quark.

For the KK-graviton analysis, as for the branon case, we need the cross-section of the subprocess $q \bar q
\rightarrow \sum_n \gamma h^{(n)}$, that was computed in \cite{GRW}. It can be written as
\begin{eqnarray}
&&\frac{d\sigma (q \bar q\rightarrow \sum_n \gamma h^{(n)} )}{dm^{2}dt}  \nonumber \\
&=&\frac{Q_{q}^{2}\alpha }{48 m^2 M_D^{2}\hat{s}^{3}tu}
\left( \frac{m^{2}\pi}{M_D^{2}} \right)^{N/2}
(\hat{s}m^{2}+4tu)(2\hat{s}m^{2}+t^{2}+u^{2})\,,
\end{eqnarray}%
where $N$ is the number of extra dimensions ($N= D- 4$), $M_D$ is associated with the fundamental
gravitational scale in the $D$-dimensional bulk space, and we have approximated the KK masses
by the continuous variable $m$.

We can see that the single photon cross section for KK-gravitons and for branons
are very similar. The continuous KK mass plays the roll
of the invariant mass of the branon pair. In fact, for $N=6$ and massless
branons, the cross section is identical with the identification:
$M_{10}^8=320\, \pi^5 f^8$ \cite{CS}. Moreover, independently of the number
of dimensions and the branon mass, the angular dependence factorizes in
the same way for branons and KK-gravitons. Therefore, it seems difficult
to distinguish between both signals by using a pseudorapidity analysis.
The invariant mass study is more promising. On the one hand, the
number of extra dimensions changes the power law dependence in the graviton
case. Therefore, this analysis can exclude the branon explanation for a possible
excess. On the other hand, a non negligible branon mass, as it would be the case
of branon dark matter, introduces a lower cut in the signature that cannot
be reproduced with the KK-graviton tower.

\begin{figure}[h]
\null\hfill\includegraphics[width=.8\textwidth]{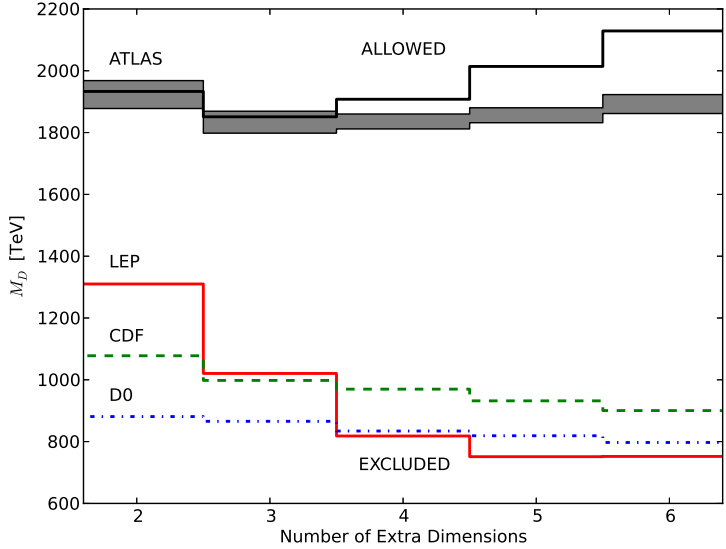}\hfill\null
\caption{Computed lowest limit (according to ATLAS data with $\sqrt{s}=7\,{\rm TeV}$ and $\int L\,dt = 4.6\,{\rm fb}^{-1}$) for the value of $M_D$ parameter of the KK--Graviton model (black solid line), versus the NLO computation of Ref.~\cite{2012arXiv1209.4625T} (gray band) and limits of LEP (solid red line), CDF (green dashed line) and D0 (blue dash dotted line).}\label{fig_Graviton}
\end{figure}

\section{Pythia simulations}\label{rafa}

To simulate the effective vertex of our $2\rightarrow 3$ processes and to estimate the experimental cuts used by ATLAS collaboration~\cite{2012arXiv1209.4625T} we have used the general purpose event generator PYTHIA 8.175~\cite{sjostrand} and its internal phase space selection machinery by inheriting the base class \emph{Sigma3Process}~\cite{2012arXiv1209.4625T}. We also have used the intrinsic random-number generator included in PYTHIA 8~\cite{Marsaglia} which, according to PYTHIA's documentation~\cite{sjostrand}, provides uniquely different random number sequences as long as the integer seeds remain below 900,000,000.

In both the branon and graviton cases, we have set $\sqrt{s}=7\,{\rm TeV}$ and the same cuts to fit the conditions of~\cite{2012arXiv1209.4625T}. The next conditions are required:
\begin{itemize}
	\item One isolated photon with $p_T>150\,{\rm GeV}$ (transverse momentum) and pseudorapidity $\lvert\eta\rvert\in [0,1.37)\cup (1.52,2.37)$.
	\item A number of jets less or equal than one. The used clustering algorithm is the anti-\emph{kT} one with a $R$ distance parameter $0.4\,{\rm GeV}$, a minimum transverse momentum $p_T > 30\,{\rm GeV}$ and a maximum pseudorapidity $\lvert\eta\rvert < 4.5$. Only observable final--state particles are included in the analysis. Both the high $p_T$ photon and the hypothetical DM particles are explicitly excluded. The true masses of particles are also used.
	\item In a cone of $\Delta R = \sqrt{(\Delta\eta)^2 + (\Delta\phi)^2}=0.4$ around the photon the sum of the energies of all the visible particles (excluded the DM particles) is less than $5\,{\rm GeV}$.
	\item A transverse missing momentum $E_T^{\rm miss}>150\,{\rm GeV}$. To compute it, we take into account all the visible particles with $\lvert\eta\rvert < 4.9$.
	\item The reconstructed photon, transverse missing momentum and jet (if found) are separated by $\Delta\phi (\gamma, E_T^{\rm miss})>0.4$, $\Delta R(\gamma,{\rm jet})>0.4$ and $\Delta\phi({\rm jet},E_T^{\rm miss})>0.4$.
	\item There are neither electrons nor positrons nor muons. This restriction applies to electrons (and positrons) with $p_T>20\,{\rm GeV}$ and $\lvert\eta\rvert < 2.47$. And to muons with $p_t > 10\,{\rm GeV}$ and $\lvert\eta\rvert < 2.4$. However, in compliance with our simulations, the effect of this restriction over the \emph{signal} is negligible although, according to Ref.~\cite{2012arXiv1209.4625T}, it is expected to reduce the background.
\end{itemize}

The internal machinery of PYTHIA 8 has been configured with the cuts (see Ref.~\cite{2012arXiv1209.4625T})
\begin{itemize}
	\item PhaseSpace:pTHatMin = 1\,GeV
	\item PhaseSpace:pTHat3Min = 1\,GeV
	\item PhaseSpace:pTHat5Min = 1\,GeV
	\item PhaseSpace:RsepMin = 0.1
\end{itemize}
The three first ones set the invariant moment $p_T$ cut in $1\,{\rm GeV}$. And the last one set the minimum separation $\Delta R = \sqrt{(\Delta\eta)^2 + (\Delta\phi)^2}$ between any two outgoing partons to $\Delta R > 0.1$.

The speed of simulation varies, so for each value of $M$ (branons) or $N$ (gravitons) we have generated 100 histograms in $E_T^{\rm miss}$ imposing that each execution take 10\,hours. The variables $f$ (branons) and $M_D$ (gravitons), since they are a multiplicative factor in the differential cross section, are introduced through a rescaling of the histogram. The only simulated hard event is our effective vertex.

We have extracted the experimental data from Ref.~\cite{2012arXiv1209.4625T}, which corresponds to the ATLAS data of 2011, with $\sqrt{s}=7\,{\rm TeV}$ and an integrated luminosity of $4.6\,{\rm fb}^{-1}$. Since the effect of an increase of both $f$ and $M_D$ is a decrease in the squared matrix element, a $\chi^2$ test is performed to find a lowest limit in both variables for each $M$ (or $N$). The \emph{experimental value} of the number of \emph{detected events} is taken as the measured events minus the background estimated by Ref.~\cite{2012arXiv1209.4625T}. And the variance $\sigma^2$, as $\sigma^2 = \sigma^2_{\rm data} + \sigma^2_{\rm background}$. A confidence limit of 95\% has been used. This variance enters into the chi--squared analysis which allows us to set lowest limits over $M_D$ (for graviton model) and $f$ (for branons).

\begin{figure}[h]
\null\hfill\includegraphics[width=.8\textwidth]{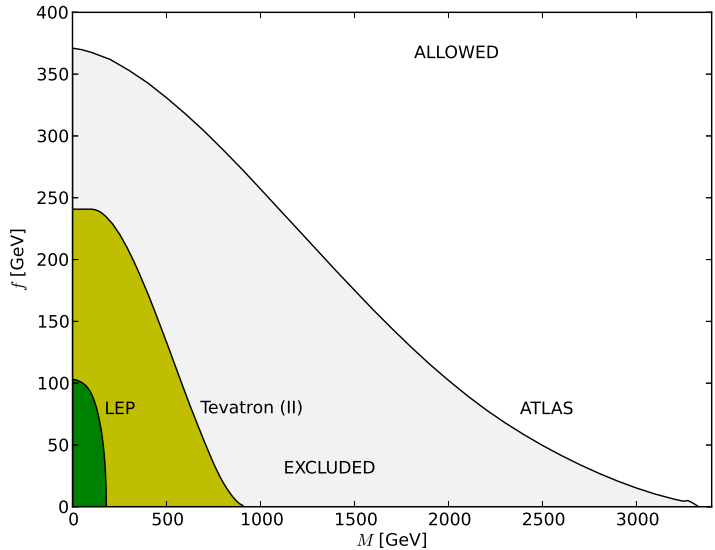}\hfill\null
\caption{Computed exclusion region (according to ATLAS data) for the value of $f$ parameter of the branon (gray area) versus the limits of the second run of Tevatron (dark yellow area, Ref.~\cite{BWHad}, $\sqrt{s}=1.96\,{\rm TeV}$ and $\int L\,dt = 200\,{\rm pb}^{-1}$) and LEP (green area, Ref.~\cite{L3Coll}, $\sqrt{s}=189-209\,{\rm GeV}$).}\label{fig_Branon}
\end{figure}

\section{Conclusions}

Our Monte Carlo computations for two models of both branon and KK--graviton are shown in Fig.~\ref{fig_MC}, along with the SM--background computation and the experimental points of ATLAS Collaboration~\cite{2012arXiv1209.4625T} ($\sqrt{s}=7\,{\rm TeV}$ and $\int L\,dt=4.6\,{\rm fb}^{-1}$). It can be seen that no signal of any of both models is found, being the experimental points compatible with the SM background.

Thus, the main goal of our computation is giving a \emph{lowest limit} in the value of $f$ parameter of the branon model for various extra dimensions $N$ (see Fig.~\ref{fig_Branon}). In all cases, a confident limit of 95\% has been used. In order to test our model, we also compute the lowest limit in the value of $M_D$ parameter of the KK-graviton model (Fig.~\ref{fig_Graviton}) and compare it with the computation of Ref.~\cite{2012arXiv1209.4625T}.

According to Fig.~\ref{fig_Branon}, although the fit is good for low values of $N$, our limit is overestimated for high $N$ by a factor $\approx 15\%$ in the worst case. However, it is remarkable that we are using a tree--level squared matrix element, while Ref.~\cite{2012arXiv1209.4625T} uses a Next to Leading Order (NLO) calculation and it has access to a full detector simulation. In any case,
our analysis provides the most constraining limits from collider experiments over the branon model.

\section{Acknowledgments}
The authors thankfully acknowledge the computer resources, technical expertise, and assistance provided by the Barcelona Supercomputing Centre ---Centro Nacional de Supercomputaci\'{o}n--- and the Tirant supercomputer support staff at Valencia. We are also pleased to thank prof. Antonio L. Maroto and prof. Torbj\"{o}rn Sj\"{o}strand for their useful help and comments. This work has been supported by MICINN (Spain) project numbers FIS 2008-01323, FIS2011-23000, FPA2011-27853-01 and Consolider-Ingenio MULTIDARK CSD2009-00064. RLD is a fellow of FPI, Ref. BES-2012-056054.


\begin{thebibliography}{99}

\bibitem{Rubakov}
%\bibitem{Lorenzana,Rubakov,Kubyshin,Gabadadze,Feruglio,Csaki,Rizzo,Burdman}
  A.~Perez-Lorenzana, AIP Conf.\ Proc.\  {\bf 562}, 53 (2001) [hep-ph/0008333];
  %%CITATION = HEP-PH/0008333;%%
  V.~A.~Rubakov,  Phys.\ Usp.\  {\bf 44}, 871 (2001) [Usp.\ Fiz.\ Nauk {\bf 171}, 913 (2001)] [hep-ph/0104152];
  %%CITATION = HEP-PH/0104152;%%
  Y.~A.~Kubyshin, hep-ph/0111027.
  %%CITATION = HEP-PH/0111027;%%
  G.~Gabadadze, hep-ph/0308112.
  %%CITATION = HEP-PH/0308112;%%
  F.~Feruglio, Eur.\ Phys.\ J.\ C {\bf 33}, S114 (2004) [hep-ph/0401033];
  %%CITATION = HEP-PH/0401033;%%
  C.~Csaki, In *Shifman, M. (ed.) et al.: From fields to strings, vol. 2* 967-1060 [hep-ph/0404096];
  %%CITATION = HEP-PH/0404096;%%
  T.~G.~Rizzo, eConf C {\bf 040802}, L013 (2004) [hep-ph/0409309];
  %%CITATION = HEP-PH/0409309;%%
  G.~Burdman, AIP Conf.\ Proc.\  {\bf 753}, 390 (2005) [hep-ph/0409322].
  %%CITATION = HEP-PH/0409322;%%

\bibitem{Rubakov:1983bb}
  V.~A.~Rubakov and M.~E.~Shaposhnikov, Phys.\ Lett.\ B {\bf 125}, 136 (1983).  %%CITATION = PHLTA,B125,136;%%

\bibitem{ADD}
  N.~Arkani-Hamed, S.~Dimopoulos, G.~R.~Dvali, Phys.\ Lett.\  {\bf B429}, 263-272 (1998) [hep-ph/9803315];
  Phys.\ Rev.\  {\bf D59}, 086004 (1999) [hep-ph/9807344];
  I.~Antoniadis, N.~Arkani-Hamed, S.~Dimopoulos and G.~R.~Dvali, Phys.\ Lett.\ B {\bf 436}, 257 (1998)
  [hep-ph/9804398]. %%CITATION = HEP-PH/9804398;%%

\bibitem{Coll}
  P. Achard {\it et al.}, Phys. Lett. {\bf B597}, 145 (2004) [hep-ex/0407017]; %%CITATION = PHLTA,B597,145;%%
  S.~Heinemeyer {\it et al.}, hep-ph/0511332; %%CITATION = ECONF,C0508141,PLEN0044;%%
  J.~A.~R.~Cembranos, A.~Rajaraman and F.~Takayama, hep-ph/0512020; %%CITATION = ECONF,C0508141,ALCPG0410;%%
  Europhys.\ Lett.\  {\bf 82}, 21001 (2008) [hep-ph/0609244]; %%CITATION = EULEE,82,21001;%%
  J.~A.~R.~Cembranos, arXiv:1301.7088 [hep-ph]. %%CITATION = ARXIV:1301.7088;%%
  A.~Juste {\it et al.}, hep-ph/0601112; %%CITATION = ECONF,C0508141,PLEN0043;%%
  ILC Collaboration, 0709.1893 [hep-ph]; %%CITATION = ARXIV:0709.1893;%%
  0712.1950 [physics.acc-ph]; %%CITATION = ARXIV:0712.1950;%%
  0712.2356 [physics.ins-det]. %%CITATION = ARXIV:0712.2356;%%

\bibitem{BWRad}
  S.~C.~Park, H.~S.~Song,  Phys.\ Lett.\  {\bf B523}, 161-164 (2001) [hep-ph/0109121];
  J.~A.~R.~Cembranos, A.~Dobado and A.~L.~Maroto, Phys.\ Rev.\ D {\bf 73}, 035008 (2006) [hep-ph/0510399]; %%CITATION = PHRVA,D73,035008;%%
  Phys.\ Rev.\ D {\bf 73}, 057303 (2006) [hep-ph/0507066]. %%CITATION = PHRVA,D73,057303;%%

\bibitem{astro}
  J.~A.~R.~Cembranos, J.~L.~Feng, A.~Rajaraman and F.~Takayama, Phys.\ Rev.\ Lett.\  {\bf 95}, 181301 (2005) [hep-ph/0507150]; %%CITATION = PRLTA,95,181301;%%
  J.~A.~R. Cembranos, J.~L.~Feng, L~E.~Strigari,  Phys.\ Rev.\  D {\bf 75}, 036004 (2007) [hep-ph/0612157]; %%CITATION = PHRVA,D75,036004;%%
  Phys.\ Rev.\ Lett.\  {\bf 99}, 191301 (2007) [0704.1658 [astro-ph]]; %%CITATION = PRLTA,99,191301;%%
  J.~A.~R.~Cembranos and L.~E.~Strigari, Phys.\ Rev.\  D {\bf 77}, 123519 (2008) [arXiv:0801.0630 [astro-ph]]; %%CITATION = PHRVA,D77,123519;%%
  J.~A.~R.~Cembranos, Phys.\ Rev.\ Lett.\  {\bf 102}, 141301 (2009) [arXiv:0809.1653 [hep-ph]]; %%CITATION = PRLTA,102,141301;%%
  Phys.\ Rev.\  D {\bf 73}, 064029 (2006) [arXiv:gr-qc/0507039]. %%CITATION = PHRVA,D73,064029;%%

\bibitem{indirect}
  J.~A.~R.~Cembranos  {\it et al.},  Phys.\ Rev.\  D {\bf 83}, 083507 (2011) [arXiv:1009.4936 [hep-ph]];  %%CITATION = PHRVA,D83,083507;%%
  Phys.\ Rev.\  D {\bf 83}, 083507 (2011) [arXiv:1009.4936 [hep-ph]];  %%CITATION = PHRVA,D83,083507;%%24
  Phys.\ Rev.\ D {\bf 85}, 043505 (2012) [arXiv:1111.4448 [astro-ph.CO]]; %%CITATION = ARXIV:1111.4448;%%
  Phys.\ Rev.\ D {\bf 86}, 103506 (2012) [arXiv:1204.0655 [hep-ph]];  %%CITATION = ARXIV:1204.0655;%%
  JCAP {\bf 1304}, 051 (2013) [arXiv:1302.6871 [astro-ph.CO]];  %%CITATION = ARXIV:1302.6871;%%
  arXiv:1305.2124 [hep-ph].  %%CITATION = ARXIV:1305.2124;%%

\bibitem{cosmo}
  T.~Biswas  {\it et al.}, Phys.\ Rev.\ Lett.\  {\bf 104}, 021601 (2010) [arXiv:0910.2274 [hep-th]];
  %%CITATION = PRLTA,104,021601;%%
  JHEP {\bf 1010}, 048 (2010) [arXiv:1005.0430 [hep-th]]; %%CITATION = JHEPA,1010,048;%%
  Phys.\ Rev.\  D {\bf 82}, 085028 (2010) [arXiv:1006.4098 [hep-th]];  %%CITATION = PHRVA,D82,085028;%%
  J.~A.~R.~Cembranos  {\it et al.}, JCAP {\bf 0907}, 025 (2009) [0905.1989 [astro-ph.CO]]; %%CITATION = JCAPA,0907,025;%%
  J.~A.~R.~Cembranos  {\it et al.}, Phys.\ Rev.\ D {\bf 86}, 021301 (2012) [arXiv:1203.6221 [astro-ph.CO]];
  %%CITATION = ARXIV:1203.6221;%% VT
  Phys.\ Rev.\ D {\bf 87}, 043523 (2013) [arXiv:1212.3201 [astro-ph.CO]];
  %%CITATION = ARXIV:1212.3201;%% VT
  F.~D.~Albareti {\it et al.}, JCAP {\bf 1212}, 020 (2012) [arXiv:1208.4201 [gr-qc]];
  %%CITATION = ARXIV:1208.4201;%%
  arXiv:1212.4781 [gr-qc]; %%CITATION = ARXIV:1212.4781;%%
  J.~A.~R.~Cembranos, Phys.\ Rev.\  D {\bf 73}, 064029 (2006) [arXiv:gr-qc/0507039]; %%CITATION = PHRVA,D73,064029;%%6
  J. A. R. Cembranos {\it et al.}, JCAP {\bf 1204}, 021 (2012) [arXiv:1201.1289 [gr-qc]];
  %%CITATION = ARXIV:1201.1289;%%
  arXiv:1109.4519 [gr-qc]. %%CITATION = ARXIV:1109.4519;%%

\bibitem{GRW}
  G.~F.~Giudice, R.~Rattazzi, J.~D.~Wells, Nucl.\ Phys.\  {\bf B544}, 3-38 (1999) [hep-ph/9811291].

\bibitem{sun}
  R.~Sundrum, Phys.\ Rev.\  {\bf D59}, 085009 (1999) [hep-ph/9805471];
  M. Bando et. al., Phys. Rev. Lett. \textbf{83}, 3601 (1999)
  T.~Kugo, K.~Yoshioka,  Nucl.\ Phys.\  {\bf B594}, 301-328 (2001) [hep-ph/9912496].

\bibitem{DoMa}
  A.~Dobado and A.~L.~Maroto,  Nucl.\ Phys.\ B {\bf 592}, 203 (2001)  [hep-ph/0007100]. %%CITATION = HEP-PH/0007100;%%

\bibitem{BW1}
  J.~A.~R.~Cembranos, A.~Dobado and A.~L.~Maroto,  Phys.\ Rev.\ Lett.\  {\bf 90}, 241301 (2003) [hep-ph/0302041]; %%CITATION = PRLTA,90,241301;%%
  Phys.\ Rev.\ D {\bf 68}, 103505 (2003) [hep-ph/0307062]. %%CITATION = PHRVA,D68,103505;%%

\bibitem{mass}
  J.~A.~R.~Cembranos, A.~Dobado and A.~L.~Maroto, Phys. Rev. {\bf D65} 026005 (2002) [hep-ph/0106322];
  %%CITATION = PHRVA,D65,026005;%%
  J.\ Phys.\ A  {\bf 40}, 6631 (2007) [hep-ph/0611024]. %%CITATION = JPAGB,A40,6631;%%

\bibitem{Alcaraz}
  J.~Alcaraz {\it et al.}, Phys. Rev.{\bf D67}, 075010 (2003) [hep-ph/0212269]. %%CITATION = PHRVA,D67,075010;%%

\bibitem{BWHad}
  J.~A.~R.~Cembranos, A.~Dobado and A.~L.~Maroto, Phys. Rev. {\bf D70}, 096001 (2004)
  [hep-ph/0405286]; %%CITATION = PHRVA,D70,096001;%%
  J.~A.~R.~Cembranos, J.~L.~Diaz-Cruz and L.~Prado, Phys.\ Rev.\ D {\bf 84}, 083522 (2011)
  [arXiv:1110.0542 [hep-ph]]. %%CITATION = ARXIV:1110.0542;%%

\bibitem{subs}
  J.~A.~R.~Cembranos, A.~Dobado and A.~L.~Maroto, Int. J. Mod. Phys. {\bf D13}, 2275 (2004) [hep-ph/0405165]; %%CITATION = IMPAE,D13,2275;%%
  A.~L.~Maroto, Phys.\ Rev.\ D {\bf 69}, 043509 (2004) [hep-ph/0310272]; %%CITATION = PHRVA,D69,043509;%%
  Phys.\ Rev.\ D {\bf 69}, 101304 (2004) [hep-ph/0402278]; %%CITATION = PHRVA,D69,101304;%%
  J.~A.~R.~Cembranos  {\it et al.}, JCAP {\bf 0810}, 039 (2008) [0803.0694 [astro-ph]]. %%CITATION = JCAPA,0810,039;%%

\bibitem{CS}
  P.~Creminelli, A.~Strumia, Nucl.\ Phys.\  {\bf B596}, 125-135 (2001) [hep-ph/0007267].

%\bibitem{2012arXiv1209.4625T}
%  The ATLAS Collaboration 2012, arXiv:1209.4625.

\bibitem{2012arXiv1209.4625T} Aad, G., Abajyan, T., 
   The ATLAS Collaboration, Phys. Rev. Lett., {\bf 110}, 011802 (2013) [arXiv:1209.4625 [hep-ex]]

\bibitem{sjostrand}
  T.~Sj\"{o}strand, S.~Mrenna and P.~Z.~Skands, JHEP {\bf 0605}, 026 (2006) [hep-ph/0603175];
  %%CITATION = HEP-PH/0603175;%%
  Comput.\ Phys.\ Commun.\  {\bf 178}, 852 (2008) [arXiv:0710.3820 [hep-ph]].
  %%CITATION = ARXIV:0710.3820;%%

\bibitem{Marsaglia}
  G.~Marsaglia, A.~Zaman and W.-W.~Tsang, Stat. Prob. Lett. {\bf 9}, 35 (1990).

\bibitem{L3Coll}
  P. Achard {\it et al.}, Phys. Lett. B {\bf 597}, 145 (2004) [hep-ex/0407017]. %%CITATION = PHLTA,B597,145;%%18

\end{thebibliography}
\end{document}